\documentclass[usenatbib,usegraphicx]{mn2e}
\usepackage{bm}
\usepackage{morefloats}
\usepackage{hyperref}
\usepackage{tabularx}
\usepackage{graphicx}
\usepackage{multirow}
\usepackage{morefloats}
\usepackage{pdflscape}

\title[No evidence of IMBHs in $\omega$ Cen and NGC 6624]
{No evidence for intermediate-mass black holes in the globular clusters $\omega$ Cen and NGC 6624}
\author[Baumgardt, He, Sweet, Drinkwater, Sollima, Hurley, Usher, Kamann, Dalgleish, Dreizler \& Husser]{H. Baumgardt$^{1}$\thanks{E-mail:
h.baumgardt@uq.edu.au}, C. He$^{1}$, S. M. Sweet$^{2,3}$, M. Drinkwater$^{1}$, A. Sollima$^{4}$, J. Hurley$^{2}$, 
\newauthor C. Usher$^{5}$, S. Kamann$^{5}$, H. Dalgleish$^{5}$, S. Dreizler$^{6}$, T.-O. Husser$^{6}$\\
$^{1}$ School of Mathematics and Physics, The University of Queensland, St. Lucia, QLD 4072, Australia \\
$^{2}$ Swinburne University of Technology, PO Box 218, Hawthorn, Victoria 3122, Australia\\
$^{3}$ ARC Centre of Excellence for All-sky Astrophysics in 3 Dimensions (ASTRO 3D)\\
$^{4}$ INAF Osservatorio Astronomico di Bologna, via Gobetti 93/3, Bologna, 40129, Italy\\
$^{5}$ Astrophysics Research Institute, Liverpool John Moores University, 146 Brownlow Hill, Liverpool L3 5RF, UK 0000-0002-7383-7106\\
$^{6}$ Institute for Astrophysics, Georg-August-University, Friedrich-Hund-Platz 1, D-37077 Göttingen, Germany\\
}

\begin{document}

\date{Accepted 2018 xx xx. Received 2018 xx xx; in original form 2018 xx xx}

\pagerange{\pageref{firstpage}--\pageref{lastpage}} \pubyear{201x}

\maketitle

\label{firstpage}

\begin{abstract}

We compare the results of a large grid of N-body simulations with the surface brightness and velocity dispersion profiles of the globular clusters
$\omega$ Cen and NGC~6624. Our models include clusters with varying stellar-mass black hole retention fractions and varying masses of a central intermediate-mass black hole (IMBH). We find that an $\sim 45,000$ M$_\odot$
IMBH, whose presence has been suggested based on the measured velocity dispersion profile of $\omega$~Cen, predicts the existence of about 20 fast-moving, $m>0.5$ M$_\odot$ main-sequence stars
with a (1D) velocity $v>60$ km/sec in the central 20 arcsec of $\omega$ Cen. However no such star is present in the HST/ACS proper motion catalogue of \citet{bellinietal2017}, strongly ruling
out the presence of a massive IMBH in the core of $\omega$~Cen.  Instead, we find that all available data can be fitted by a model that contains 4.6\% of the mass of $\omega$ Cen in a centrally concentrated cluster of stellar-mass black holes. 
We show that this mass fraction in stellar-mass BHs is compatible with the predictions of stellar evolution models of massive stars.

We also compare our grid of $N$-body simulations with NGC 6624, a cluster recently claimed to harbor a $20,000$ M$_\odot$ black hole
based on timing observations of millisecond pulsars. However, we find that models with $M_{IMBH}>1,000$ M$_\odot$ IMBHs
are incompatible with the observed velocity dispersion and surface brightness profile of NGC 6624,ruling out the presence of a massive IMBH in this cluster.
Models without an IMBH provide again an excellent fit to NGC 6624.
\end{abstract}
 

\section{Introduction} \label{sec:intro}

Black holes were long considered to be a mathematical curiosity, but nowadays their existence has firm observational support.
Until recently, observational evidence for black holes has mainly been gathered in two distinct mass ranges: stellar mass black holes,
which are produced as the end product of the stellar evolution of massive stars \citep{f99}, and supermassive black holes with masses
10$^6$-10$^{10}$ M$_\odot$, which are found in the centres of galaxies \citep{getal00, gultekinetal2009}.

In recent years, evidence has also been accumulating for the existence of intermediate-mass black holes (IMBHs) with masses in the range 10$^3$-10$^5$ M$_\odot$. 
First, some IMBHs have been found in the centres of dwarf galaxies. \citet{betal04} for example found a $10^5$ M$_\odot$ black hole at the centre
of the Seyfert 1 galaxy POX 52 through optical imaging and stellar radial velocity measurements.
\citet{farrelletal2009} found evidence that the ultraluminous X-ray source in the galaxy ESO243-49
is powered by an accreting black hole with mass $10^2$ to $10^5$ M$_\odot$. Further evidence for an IMBH nature of the
accreting black hole was later found by \citet{webbetal2010} and \citet{servillatetal2011}. 
Accreting IMBH candidates were also found at the centres of the galaxies NGC 404 \citep{nylandetal2012} and NGC 3319 \citep{jiangetal2018}, making it plausible
that IMBHs could be intermediate steps in the formation of supermassive black holes.
Most recently \citet{linetal2018} found that a luminous X-ray outburst in a massive star cluster near the lenticular galaxy 6dFGS gJ215022.2-055059
was most likely powered by the tidal disruption of a star by a 50,000 M$_\odot$ IMBH.

IMBHs might also exist in globular clusters, created through either the formation of a central cluster of compact remnants which later merge due to the emission of gravitational waves
\citep{mh02,mt02}, run-away merging of massive main sequence stars
within the first few Myrs after cluster formation \citep{pzetal2004},
or the repeated formation of tight binaries between a stellar mass black hole and main sequence stars followed by mass accretion
onto the black hole and subsequent growth of the black hole over longer timescales \citep{gierszetal2015}. IMBHs might also be
the remnants of $\sim 10^4$ M$_\odot$ supermassive stars that have been suggested as the sources of the observed abundance anomalies in globular clusters
\citep{denissenkovhartwick2014}.

Observational evidence for the existence of IMBHs has been reported in about 20 Galactic globular clusters
based on either stellar kinematics \citep[e.g.][]{gerssenetal2002}, X-ray or radio signals from accretion of interstellar gas \citep{ulvestadgreeneho2007} or 
the acceleration of pulsars \citep{kiziltanetal2017,pereraetal2017a}. In particular, \citet{noyolaetal2008}, \citet{jalalietal2012} and \citet{baumgardt2017} found evidence for a
$40,000$ M$_\odot$ IMBH in the centre of $\omega$ Cen based on the velocity dispersion and surface brightness profile of this cluster.
Since $\omega$ Cen is thought to be the nuclear cluster of a tidally disrupted dwarf galaxy \citep[e.g.][]{bekkinorris2006}, such a discovery could provide a link between IMBHs and supermassive
black holes. However the presence of an IMBH in $\omega$ Cen was challenged by \citet{vdmanderson2010}, who created models of $\omega$ Cen that
fitted the velocity dispersion profile of the cluster without the need for
an IMBH, and \citet{zocchietal2019} who fitted the velocity dispersion profile of $\omega$ Cen by a model that contained a centrally concentrated
cluster of stellar-mass black holes. Furthermore, \citet{haggardetal2013} found no evidence of radio signals from gas accretion onto a central black hole in $\omega$ Cen.

In addition, \citet{pereraetal2017a} and \citet{pereraetal2017b} found evidence for a massive IMBH in NGC~6624 based on timing observations of
several pulsars close to the cluster centre. However \citet{gielesetal2018} were able to explain the observed period changes by a cluster model that did
not contain an IMBH. In summary, there is currently 
no undisputed case for an IMBH in any Galactic globular cluster. If IMBHs exist in globular clusters, most of them must have masses
of less than a few thousand M$_\odot$, otherwise their influence on the velocity dispersion profiles \citep{baumgardt2017} or radio emission from
the accretion of interstellar gas \citep{straderetal2012,tremouetal2018} should have been detected.

In the present paper we use theoretical models to investigate whether the surface brightness and velocity dispersion profiles of the globular clusters $\omega$ Cen and NGC~6624 require
the presence of IMBHs in these clusters. Our models are based on direct $N$-body simulations, which follow the evolution of both clusters under
the combined influence of stellar evolution and two-body relaxation. For $\omega$ Cen we also investigate a centrally concentrated cluster of stellar mass black holes as an
alternative to an IMBH. Our paper is organised as follows: In Section~2 we describe the observational data used in this work. In Section~3 we describe
the grid of $N$-body simulations used to fit the velocity and surface brightness profiles and
the stellar mass functions of globular clusters. In Section~4 we compare the $N$-body models with the observations and we draw our conclusions in Section~5.

\section{Observational data}

\subsection{$\omega$ Cen}

Our main source for the kinematic data on $\omega$ Cen are the radial velocity dispersion profiles recently published by \citet{baumgardt2017} and \citet{baumgardthilker2018}.
\citet{baumgardt2017} calculated the velocity dispersion based on $\sim 4,500$ individual stellar radial velocities from published literature data,
while \citet{baumgardthilker2018} determined the radial velocities of an additional 1,000 cluster stars from unpublished ESO/FLAMES spectra. 
In order to improve the coverage of the outer regions of $\omega$ Cen, we added to this data a set of 10 AAOmega/2dF observations of $\omega$ Cen made between
July 2007 and May 2011, that we downloaded from the AAT Data Archive.
We restricted ourselves to AAOmega spectra taken with the 1700D grating which have a spectral resolution of $R=10,000$, the highest of all available AAOmega gratings. 
The basic data reduction of these spectra 
was done with the program {\tt 2dfdr}, which also performed the heliocentric correction of the spectra. We calculated radial velocities from the reduced spectra
with the help of the IRAF task {\tt fxcor}, which is based on the Fourier cross-correlation method developed by \citet{tonrydavis1979}.
For the cross-correlation, we used as template the spectrum of a cool giant star that we created with the help of the stellar synthesis
program {\tt SPECTRUM} \citep{graycorbally1994} using {\tt ATLAS9} stellar model atmospheres \citep{castellikurucz2004} with a metallicity of $[Fe/H]=-1.50$ as input.

In total we obtained 6,500 radial velocities of stars in the field of $\omega$ Cen, from which we calculated the velocity dispersion profile of $\omega$ Cen using
a maximum-likelihood approach: We first cross-correlated the different data sets against each other to bring them to a common mean radial velocity
and cross-matched the stellar positions against the {\tt Gaia} DR2 catalogue. We next removed all stars that have proper motions incompatible with the mean cluster
motion determined by \citet{baumgardtetal2019}. The mean cluster
velocity and velocity dispersion profile were then determined using all remaining stars and the membership probability of each star was determined based on the
velocity dispersion of the cluster. We then removed stars with radial velocities differing by more then 3$\sigma$ from the cluster mean from the sample and 
calculated a new mean cluster velocity and velocity dispersion profile. This procedure was repeated until a stable solution for the list of cluster members 
and the velocity dispersion profile was found and we reached this convergence within two or three steps.

In order to increase the coverage of the central cluster region, we also use the velocity dispersion profile published
by \citet{kamannetal2018} based on VLT/MUSE observations in our modeling. We accompany the line-of-sight radial velocity data with the HST based proper motion
dispersion profile of \citet{watkinsetal2015a} in the inner cluster parts and the {\tt Gaia} DR2 proper motion dispersion profile from \citet{baumgardtetal2019}. 
Finally, we use the
catalogue of $240,000$ stars with measured HST proper motions and photometry derived by \citet{bellinietal2014} and \citet{bellinietal2017}. When transforming the proper motions into velocities,
we assume a distance of $d=5.24$ kpc to $\omega$ Cen \citep{baumgardtetal2019}. The resulting velocity dispersion profile of $\omega$ Cen is shown in Fig.~\ref{ocen1}. It can
be seen that the velocity dispersion is roughly constant in the central 100'', and decreases further out before leveling off beyond about 1000''. There is generally very good agreement between the HST proper motion based velocity dispersion profile 
and the line-of-sight radial velocities. Inside 10'' both the line-of-sight radial velocity dispersion profile as well as the proper motion dispersion profile
show some larger scatter due to a lack of stars with measured kinematics.

In addition to the velocity dispersion profile, we also fit the observed surface brightness profiles with our $N$-body models. 
For both $\omega$ Cen and NGC~6624, we create surface brightness profiles by combing the HST based surface brightness profile of \citet{noyolagebhardt2006} in the inner cluster
parts with the ground-based data of \citet{trageretal1995} at larger radii. The mass function of $\omega$ Cen has been measured by 
\citet{sollimaetal2007}, who found a steep, Salpeter-like increase of the mass function between 0.5 M$_\odot <m<0.8$ M$_\odot$, a break in the mass function at $m=0.5$ M$_\odot$ 
and a flatter increase below this mass. Since this mass function is close to a Kroupa initial mass function, we use $N$-body models with a Kroupa mass function to model $\omega$ Cen.
A mass function rich in low-mass stars is also reasonable given the
high mass and long relaxation time of $\omega$ Cen, which means that only little mass segregation and little dynamical mass loss have occurred over a Hubble time.

\subsection{NGC 6624}

Radial velocities for 19 stars in the centre of NGC 6624 were determined by \citet{pryoretal1991}. In addition, \citet{baumgardthilker2018} determined the radial velocities of 
125 stars in the field of NGC 6624 based on archival VLT/FLAMES observations (proposal ID 083.D-0798(D), PI B. Lanzoni) and another 8 cluster stars from Keck/NIRSPEC observations
(Keck proposal ID U17NS, PI: M. Rich). We add to this data set, 58 stars in the central region with measured radial velocities from the WAGGS survey \citep{usheretal2017}. The 
observations were made using the WiFeS integral field spectrograph \citep{dopitaetal2007,dopitaetal2010} and the basic data reduction was done as described in \citep{usheretal2017}. 
Using \textsc{PampelMuse} \citep{kamannetal2013}, stellar spectra were extracted from the WAGGS datacubes and radial velocities were determined with the IRAF task {\tt fxcor}.
Further details will be described in a forthcoming paper (Dalgleish et al. in prep). We furthermore added radial velocities based on MUSE observations of NGC~6624
that were taken during the nights of 2015-05-11 and 2017-10-17, as part of observing programmes 095.D-0629 and 0100.D-0161 (PI: Dreizler). 
The reduction and analysis of the data were performed as described in \citet{kamannetal2018}. In particular, we used \textsc{PampelMuse} 
to extract stellar spectra from the reduced data cubes, while the derivation of the final radial velocities was done with \textsc{spexxy} 
\citep[see][]{husseretal2016}. For the present study, we selected a high-quality sample of 241 stars with $V<17$ and radial velocity uncertainties $<1.5\,{\rm km\;s^{-1}}$ 
from the full MUSE sample for NGC~6624.

In order to increase the number of stars with measured radial velocities in the outer cluster parts, we also observed NGC~6624 for one 
half-night using the {\tt DEIMOS} spectograph \citep{faberetal2003} on the Keck II telescope (Programm ID: Z252, PI M. Drinkwater). Observations were 
performed on 19 July 2017 with 0.6" seeing and some thin cirrus, using the 
1200G grating with a central wavelength of 8000 $\AA$ and the OG550 order-blocking filter. Four slitmasks were observed at position angles (PAs) of 0, 90, 270 and 
315 degrees, in order to maximise the number of stars near the centre of the cluster, where the proportion of member stars is expected to be higher, and to maximise 
the spatial coverage of the outer regions. Slits were placed on a total of 685 unique targets, comprised of 545 stars with 2MASS 
coordinates and 140 with positions derived from HST imaging. Each mask had seven stars in common with the VLT/FLAMES data set to assist 
in radial velocity calibration. Exposure times were 3 x 800 seconds for the mask with PA = 0, 3 x 850 s for PA = 270, and 3 x 900 s for the masks with PA = 90 and 315. 

The {\tt DEIMOS} spectra were reduced with the help of the {\tt DEEP2} data reduction pipeline developed by the {\tt DEEP2} survey team 
\citep{cooperetal2012, newmanetal2013}. Individual stellar radial velocities were again determined from the reduced spectra with the help of the 
IRAF {\tt fxcor} task.
To correct for residual systematic errors in the absolute wavelength calibration of the {\tt DEIMOS} spectra, we cross-correlated them against a telluric template
spectrum that was kindly provided to us by Tony Sohn and Emily Cunningham. Since the telluric lines should be at zero radial velocity, wavelength calibration 
errors can be corrected from the radial velocity of these lines. Final radial velocities for each star were then calculated according to $v_r=v_{obs}-v_{tel}-v_{hel}$, where $v_{obs}$ 
is the radial velocity derived from the stellar template, $v_{tel}$ the radial velocity from the telluric spectrum and $v_{hel}$ the heliocentric
correction. In total we were able to determine the radial velocities of 264 stars from the {\tt DEIMOS} spectra. Table~\ref{indveltab} gives the individual radial velocities
that we have derived from the {\tt DEIMOS} spectra. The membership probabilities in Table~\ref{indveltab} are calculated as described in \citet{baumgardthilker2018}.

Our final data set consists of about 600 stars with measured radial velocities in the field of NGC 6624, out of which about 200 stars are cluster members.
35 stars have measured radial velocities from both the VLT/FLAMES observations and our DEIMOS observations and 31 stars are in common between the MUSE data and the combined 
VLT/FLAMES and Keck/DEIMOS data set. Virtually all the stars measured by Pryor et al. as well as the stars measured by the WAGGS survey are
also in the MUSE sample. The good overlap between the different data sets allows us to bring them to within 0.3 km/sec of each other. Since the remaining uncertainty 
adds quadratically to the true velocity dispersion profile, it does not significantly influence our measurement of the final velocity dispersion profile.

In order to calculate the velocity dispersion profile of NGC~6624 from the individual stellar radial velocities, we again cross-correlated the different
data sets against each other to bring them to a common mean radial velocity and then selected as possible cluster members all stars with radial velocities between 
$30$ km/sec $<v<80$ km/sec and {\tt Gaia} DR2 proper motions that match the mean cluster proper motion
determined by \citet{baumgardtetal2019}.  Due to the significant stellar background density, we restricted 
the member search to distances less than 200'' from the cluster centre, since outside this radius a reliable membership determination was not possible. The calculation of the velocity dispersion profile
was again done via a maximum-likelihood approach, following the procedure described above for $\omega$ Cen.  
The resulting velocity dispersion profile is presented in Table~1.

We also used the proper motion velocity dispersion profile of \citet{watkinsetal2015a} who determined the velocity dispersion profile inside 80'' based on $\sim1,800$ stars
with magnitudes brighter than about 1.5 mag below the main-sequence turn-off.
We finally used the stellar mass function of NGC~6624 measured by \citet{saracinoetal2016} in the central 40'' from ultra-deep, adaptive optics
assisted $J$ and $K_S$ band {\tt Gemini/GSAOI} images to constrain the mass function of the best-fitting $N$-body models.
\begin{table}
\caption{Observed line-of sight velocity dispersion profiles of $\omega$ Cen and NGC~6624. For each bin, the table gives the name of the cluster, the number of stars used to calculate the radial velocity dispersion, the average distance of stars from the cluster centre, and the velocity dispersion together with the $1\sigma$ upper and lower error bars.}
\begin{tabular}{lrr@{.}lcc@{}c}
\hline
 & \\[-0.3cm]
Cluster  & \multirow{2}{*}{$N_{RV}$} & \multicolumn{2}{c}{$r$} & \multicolumn{1}{c}{$\sigma$} & \multicolumn{1}{c}{$\Delta \sigma_u$} &  \multicolumn{1}{c}{$\Delta \sigma_l$}\\
 & & \multicolumn{2}{c}{[arcsec]} & \multicolumn{1}{c}{[km/sec]} & \multicolumn{1}{c}{[km/sec]} & \multicolumn{1}{c}{[km/sec]}\\[+0.1cm]
\hline
 & \\[-0.3cm]
$\omega$ Cen &  95 &   44 & 10 & 19.09 & 1.46 & 1.29 \\
$\omega$ Cen & 100 &   75 & 11 & 19.00 & 1.44 & 1.29 \\
$\omega$ Cen & 195 &  101 & 14 & 15.72 & 0.84 & 0.77 \\
$\omega$ Cen & 195 &  126 & 99 & 17.94 & 0.95 & 0.87 \\
$\omega$ Cen & 195 &  148 & 85 & 15.89 & 0.85 & 0.78 \\
$\omega$ Cen & 195 &  171 & 92 & 14.63 & 0.78 & 0.71 \\
$\omega$ Cen & 195 &  198 & 68 & 15.39 & 0.82 & 0.76 \\
$\omega$ Cen & 195 &  222 & 19 & 14.44 & 0.77 & 0.70 \\
$\omega$ Cen & 195 &  244 & 66 & 14.30 & 0.76 & 0.70 \\
$\omega$ Cen & 195 &  266 & 18 & 12.84 & 0.68 & 0.63 \\
$\omega$ Cen & 195 &  288 & 99 & 13.83 & 0.73 & 0.68 \\
$\omega$ Cen & 195 &  315 & 19 & 12.89 & 0.69 & 0.63 \\
$\omega$ Cen & 195 &  341 & 95 & 13.27 & 0.70 & 0.65 \\
$\omega$ Cen & 195 &  378 & 73 & 11.72 & 0.62 & 0.58 \\
$\omega$ Cen & 195 &  426 & 96 & 11.96 & 0.64 & 0.59 \\
$\omega$ Cen & 195 &  480 & 76 & 12.66 & 0.67 & 0.62 \\
$\omega$ Cen & 195 &  536 & 14 & 11.22 & 0.60 & 0.55 \\
$\omega$ Cen & 195 &  590 & 69 & 10.86 & 0.58 & 0.54 \\
$\omega$ Cen & 195 &  681 & 63 & $\;$9.58 & 0.51 & 0.47 \\
$\omega$ Cen & 195 &  943 & 29 & $\;$9.52 & 0.51 & 0.47 \\
$\omega$ Cen & 195 & 1279 & 52 & $\;$8.32 & 0.45 & 0.41 \\
$\omega$ Cen & 120 & 1898 & 97 & $\;$6.77 & 0.47 & 0.42 \\
 & \\[-0.3cm]
NGC 6624 & 46 &  4 & 33 & $\;$6.41 & 0.75 & 0.63 \\
NGC 6624 & 46 &  9 & 93 & $\;$7.13 & 0.83 & 0.71 \\
NGC 6624 & 46 & 15 & 15 & $\;$6.15 & 0.72 & 0.61 \\
NGC 6624 & 46 & 20 & 27 & $\;$6.04 & 0.71 & 0.60 \\
NGC 6624 & 46 & 26 & 54 & $\;$5.11 & 0.60 & 0.51 \\
NGC 6624 & 46 & 40 & 96 & $\;$5.15 & 0.61 & 0.52 \\
NGC 6624 & 50 & 95 & 83 & $\;$3.25 & 0.39 & 0.32 \\
 & \\[-0.3cm]
\hline
\end{tabular}
\label{veldistab}
\end{table}

\section{$N$-body models}

We used the grid of $N$-body simulations presented by \citet{baumgardt2017} and \citet{baumgardthilker2018} to model $\omega$ Cen and NGC~6624 and to
derive limits on the presence of intermediate-mass black holes in these clusters. \citet{baumgardt2017} and \citet{baumgardthilker2018} have
run a grid of 1400 $N$-body simulations of star clusters containing $N=100,000$ or $N=200,000$ stars using NBODY6 \citep{aarseth1999,nitadoriaarseth2012}, varying the initial density profile and 
half-mass radius, the initial mass function, the cluster metallicity and the mass fraction of an intermediate mass black hole in the clusters. All models
consisted initially only of single stars and formed binaries only through encounters of stars in the cluster centres. Given the low observed
binary fraction in globular clusters (only of order 10\% see e.g. \citet{miloneetal2012} and \citet{jibregman2013})
we do not think that our results would significantly change with the inclusion of primordial binaries.
The basic strategy that we use to compare the $N$-body models with the observed surface brightness and velocity dispersion profile of a globular
cluster is the same as in these two papers and we refer the reader to these papers for a detailed description. In short, the $N$-body simulations
were run up to an age of $T=13.5$ Gyr and final cluster models were calculated by taking 10 snapshots from the simulations 
centered around the age of each globular cluster. The combined snapshots of the $N$-body clusters were then scaled in mass and radius to match
the density and velocity dispersion profiles of the observed globular clusters and the best-fitting model was determined from an interpolation
in the grid of $N$-body models. When comparing with the $N$-body data, we assume an age of $T=11.25$ Gyr
for NGC~6624 \citep{vandenbergetal2013} while for $\omega$ Cen
we assume an age of $T=12.0$ Gyr. Our results are however not very sensitive to the adopted cluster age.

In order to model $\omega$ Cen, we ran an additional grid of models in which we varied the retention fraction of stellar-mass black holes.
The simulations by \citet{baumgardt2017} and \citet{baumgardthilker2018} assume a retention fraction of 10\% for the black holes that
form in the simulations, with the remaining black holes given such high kick velocities upon formation that they immediately leave the star clusters. 
Such a retention fraction could be too small for $\omega$ Cen, since the models by \citet{baumgardthilker2018} predict a central escape velocity of $v_{esc}=63$ km/sec
for $\omega$ Cen, which is one of the highest central escape velocities of all Galactic globular clusters. Given that the initial escape velocity was probably even higher
due to stellar evolution driven mass loss and cluster expansion, a significant fraction of black holes could have been retained in $\omega$ Cen.
In this paper we therefore ran additional simulations of star clusters without IMBHs, but with stellar-mass black hole retention fractions of 30\%, 50\% and 
100\%. The initial mass function of stars in these simulations was also assumed to be a \citet{kroupa2001} mass function initially. 


We also ran additional simulations for NGC~6624 since the IMBH models of \citet{baumgardt2017} only contain IMBHs with up to 2\% of
the cluster mass at $T=12$ Gyr, while the $20,000$ M$_\odot$ IMBH inferred by \citet{pereraetal2017a,pereraetal2017b} implies a much larger mass fraction.
We therefore extended the grid of IMBH models of \citet{baumgardt2017} to also contain IMBH masses of 5\% and 10\% of the final cluster mass.

\section{Results}

\subsection{$\omega$ Cen}

We start our discussion of $\omega$ Cen by comparing the best-fitting $N$-body models with and without an IMBH to the observed surface brightness and velocity
dispersion profile of $\omega$~Cen. Three sets of models were calculated. In the first set of models, we kept the black hole retention fraction 
fixed at 10\% and varied only the initial cluster radius and the initial surface brightness profile, quantified by the King concentration parameter $c$. 
This was done until we found the best-fitting model to the observed surface brightness and velocity dispersion profile. These models are the same models as the $N$-body models used by 
\citet{baumgardt2017} and \citet{baumgardthilker2018}. When scaled to $\omega$ Cen, these models produce about 34,000 M$_\odot$ in stellar mass black holes 
after stellar evolution and velocity kicks have been applied, out of which $14,000$ M$_\odot$ remain in the cluster by $T=12$ Gyr.
In the second set of models we varied the assumed black hole retention fraction in addition to the
the initial cluster radius $r_h$ and initial surface density profile.
In the third set of models we fixed the retention fraction of stellar-mass black holes to 10\%, but varied the mass fraction of a central IMBH from 0.5\% to 2\% of the final cluster mass, corresponding in the case of $\omega$ Cen
to IMBHs with masses between about 12,000 to 50,000 M$_\odot$, to find the best fit to the
observed surface brightness and velocity dispersion profile. 
\begin{figure*}
\begin{center}
\includegraphics[width=0.99\textwidth]{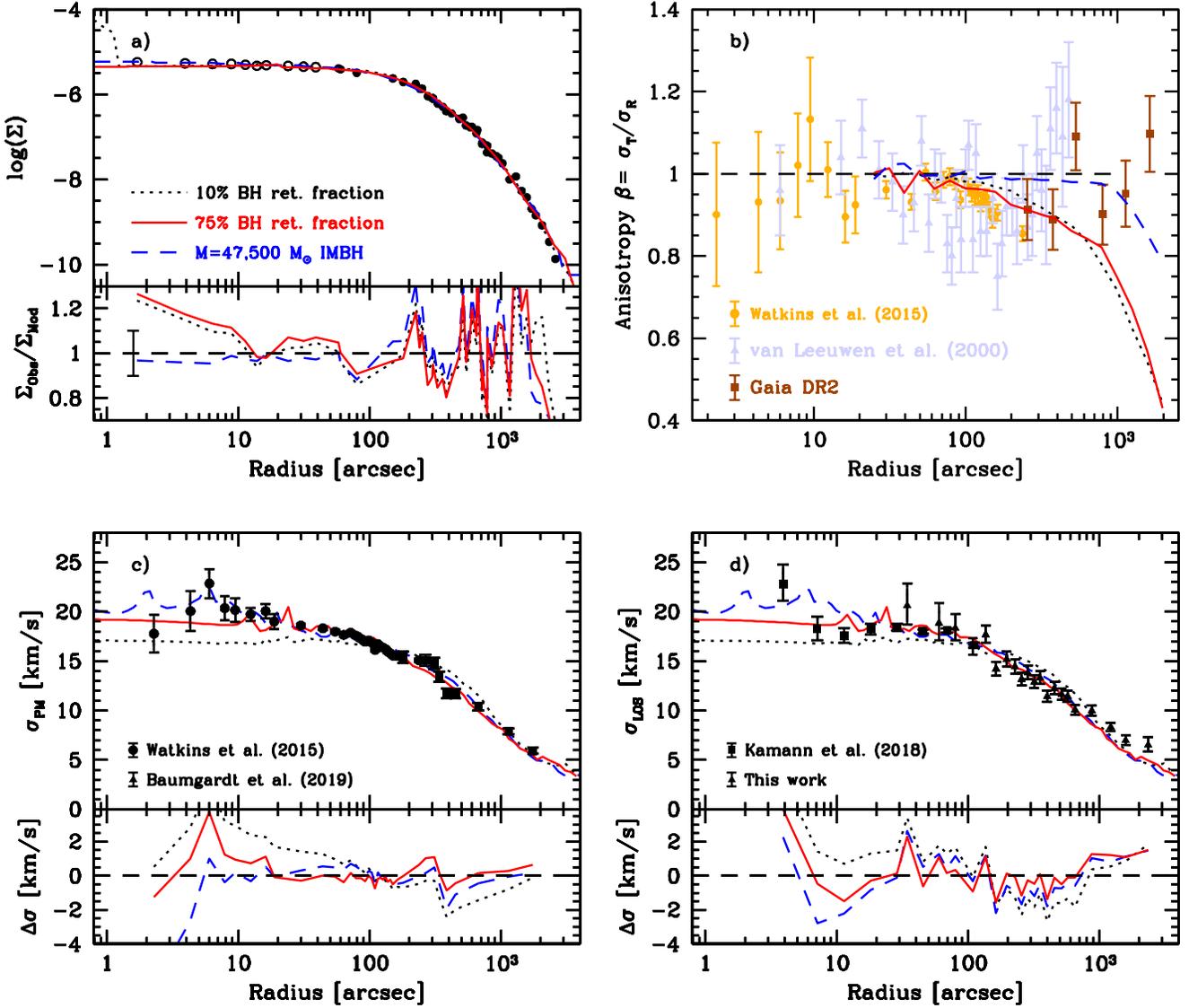}
\end{center}
\caption{Fit of the surface brightness profile (panel a) and the velocity dispersion profile (panels c and d) of $\omega$ Cen for the best-fitting $N$-body models.
 Upper panels show the actual profiles, lower panels show the differences between the observed and modeled profiles. The surface brightness profile from \citet{trageretal1995} is
 shown by solid circles while open circles show the data from \citet{noyolagebhardt2006}. The errorbar in the lower left of panel a) depicts
  an uncertainty of 0.1 mag. In panels c) and d), the observed velocity dispersions are from \citet{watkinsetal2015a} (circles), \citet{kamannetal2018} (squares) and
this work (triangles).
 Shown are the best-fitting IMBH model (blue dashed lines) and the best fitting no IMBH model with a retention fraction of stellar-mass black holes of 75\%
  (red solid lines). Also shown is the best-fitting model with a 10\% retention fraction of black holes (black dotted lines). All three models fit the
  surface brightness profile within the observational uncertainties outside the central 10''. The model with a low assumed retention fraction of stellar-mass black holes has too
  little mass in the centre and underpredicts the observed velocity dispersion in the centre and overpredicts it at larger radii. The model with a high stellar-mass black hole
   retention fraction provides a significantly better fit. Panel b) compares the anisotropy profile of $\omega$ Cen with all three models. The models are in agreement with the
  observed profile out to several hundred arcsec.}
\label{ocen1}
\end{figure*}

Panel a) of Fig.~\ref{ocen1} shows the resulting fits of the surface brightness and velocity dispersion profile of $\omega$ Cen.
Based on the differences between the polynomial fit of the surface brightness profile by \citet{trageretal1995} and the actual surface brightness
measurements, we estimate a typical uncertainty
of the observed surface brightness of $\omega$ Cen of $\Delta \Sigma = 0.1$ mag. This uncertainty is shown in the lower left of panel a) in Fig.~\ref{ocen1}. It can be seen that 
all models provide fits of similar accuracy to the surface brightness profile of $\omega$ Cen, at least outside the central 10''. Inside this radius the IMBH model provides a better fit to the
weak cusp which \citet{noyolagebhardt2006} found in the surface brightness profile. We note however that several determinations of the centre of $\omega$ Cen exist and that
\citet{andersonvandermarel2010} did not find a rise in the surface brightness profile of $\omega$ Cen around their centre. We conclude that the surface brightness profile alone 
cannot be used to discriminate between
the different models. The main reason for this is the low central concentration and long relaxation time of $\omega$ Cen, which means that the cluster 
is still far from core collapse and has not undergone much dynamical evolution. Such evolution would be necessary to establish a weak cusp profile in the surface brightness profile,
which would separate models with and without IMBHs from each other \citep[see discussion in][]{baumgardtetal2005}.

Panels c) and d) of Fig.~\ref{ocen1} compare the velocity dispersion profiles predicted by the different models against the proper motion and radial velocity dispersion profile of $\omega$ Cen.
It can be seen that the model with an initial retention fraction of stellar mass black holes of 10\% (about 14,000 M$_\odot$ in stellar mass black holes after $T=12$ Gyr) underestimates the velocity 
dispersion in the central 100'' by about 2 km/sec. It also overpredicts the velocity dispersion profile beyond 200'' and can therefore be rejected.
The best-fitting model in which the black hole retention fraction and the total mass in stellar mass black holes was left as a free parameter is shown by
a solid, red line in Fig.~\ref{ocen1}. For an initial retention fraction of stellar-mass black holes of 75\% $\pm$ 8\%, (corresponding to 165,000 M$_\odot$ or about 4.6\% $\pm$ 0.5\%
of the cluster mass in stellar-mass black holes at $T=12$ Gyr), this model
provides a significantly better fit to the velocity dispersion profile, especially in the inner cluster parts. Due to mass segregation of the stellar mass black holes into the centre,
the mass in the centre is increased compared to the model with a low retention fraction and this increases the velocity dispersion of the stars. Similarly, a model with a $M_{IMBH}=47,500$
M$_\odot$ IMBH provides a good fit to the velocity dispersion profile. Our results confirm the models of \citet{zocchietal2019} who already found that a dense cluster
of stellar-mass black holes containing 5\% of the total cluster mass can mimic the influence of an intermediate-mass black hole. 

\subsubsection{Central velocity distribution}

\begin{figure}
\begin{center}
\includegraphics[width=0.95\columnwidth]{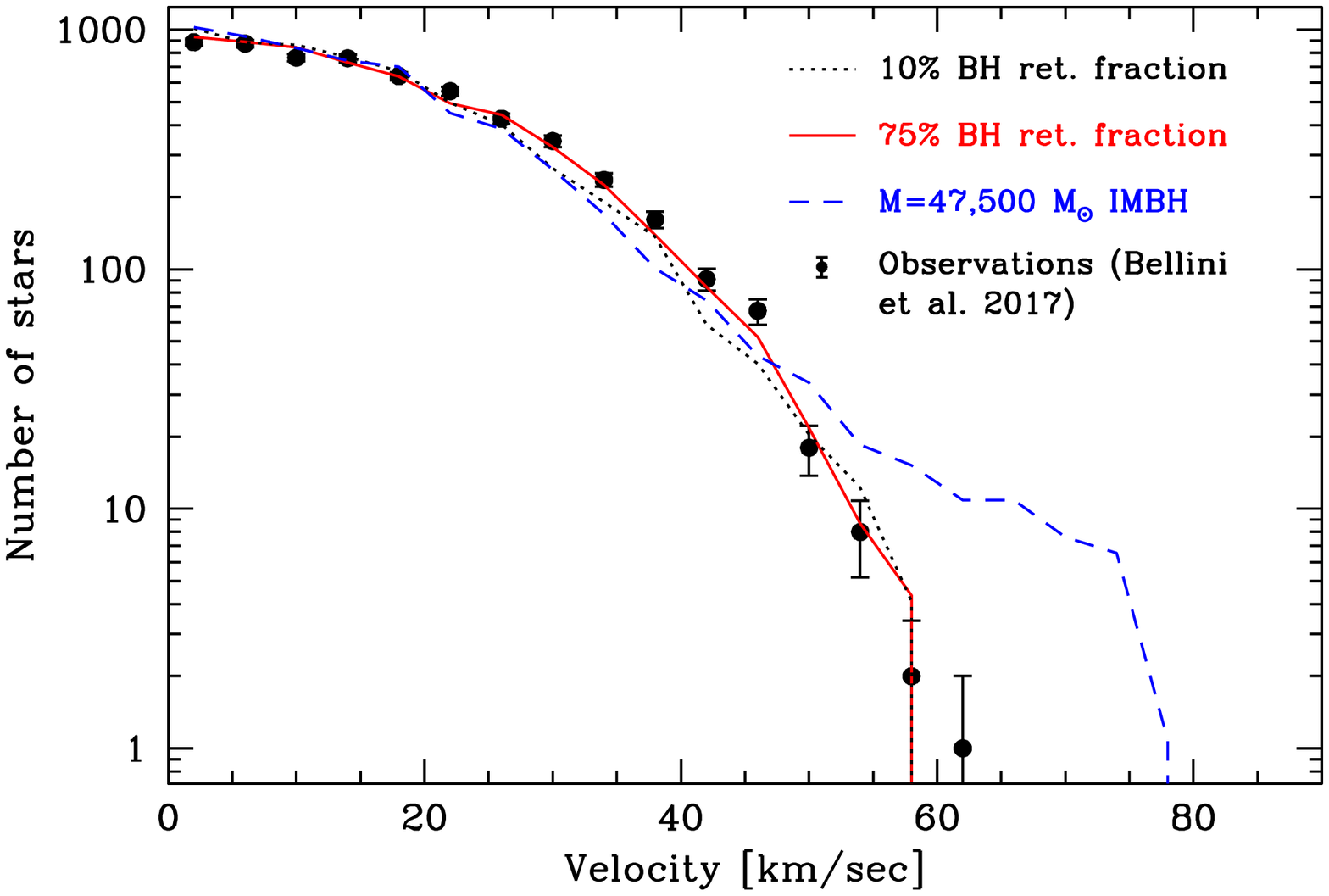}
\end{center}
\caption{Velocity distribution of stars within 20'' of the centre of $\omega$ Cen. Shown is the 1D velocity distribution for the stars with measured proper motions by
\citet{bellinietal2017} (black dots) and the best-fitting $N$-body model with an IMBH (blue dashed line) and with 10\% and 75\% BH retention fractions (black dotted and red solid lines). The model with
a high retention fraction of stellar-mass black holes provides the best fit to the observed distribution. The IMBH model predicts about 20 high-velocity stars
with $v>62$ km/sec while none is seen in the observations. The velocity distribution of the IMBH and low black hole retention fraction models also have the wrong shape 
for stars moving with less than 60 km/sec.}
\label{ocenveldis}
\end{figure}

Fig. \ref{ocenveldis} shows the 1D velocity distribution in the centre of $\omega$ Cen for the different $N$-body models and compares them with the 
observed velocity distribution which we calculated from the proper motion data published by \citet{bellinietal2017}. For the observed profile, we used
all stars within a projected radius of 20 arcsec of the cluster centre as determined by \citet{goldsburyetal2013}. This area is large enough to encompass 
the cluster centres of \citet{noyolagebhardt2006} and \citet{vdmanderson2010} as well, which are about 12'' and 0.5'' respectively away from the
Goldsbury et al. cluster centre. We restrict ourselves to stars that have reduced
$\chi_r^2$ values of less than 1.5, have a $N_{Used}/N_{found}$ ratio between the number of data points used for PM fits $N_{Used}$ to
the number of data points available $N_{found}$ of larger than 0.85, proper motion errors of less than 5~km/sec, and have velocities within 100 km/sec 
of the mean cluster velocity. For the $N$-body models
we use all main-sequence and giant stars that are more massive than 0.5 M$_\odot$ to roughly cover the same mass range as the stars with observed proper motions in the
Bellini et al. sample. We scale all theoretical distributions to contain the same number of stars as are in the observed sample. We also add random velocity errors that 
follow a Gaussian with a width of 3 km/sec to the stars from the $N$-body simulations to mimic the influence of velocity errors in the observations.

It can be seen that the model without an IMBH but a high black hole retention fraction matches the observed velocity distribution very well. In both data sets the fastest
star is moving with about 62 km/sec and the overall shape of the observed velocity distribution is also matched very well. A Kolmogorov-Smirnov (KS) test between the theoretical and observed data gives
a 15\% chance that both distributions are drawn from the same underlying distribution. Given the considerable uncertainties in e.g. the mass distribution of formed black holes,
and the large number of observed stars in Fig.~1, which makes modeling their exact distribution challenging, we consider the black hole models to be in very good agreement with the observations. 
In contrast, the best-fitting IMBH model leads to a significantly less satisfactory fit of the velocity distribution. The stellar distribution
extends to too high velocities, the IMBH model predicts 20 stars with velocity $v>62$ km/sec while none is seen in the observations. The reason for the larger number of stars
with very high velocities is the lowering of the central potential well due to the IMBH, which is more effective than that caused by a more widely distributed population of
stellar-mass black holes. The absence of fast moving stars in the observations cannot be a selection effect since such stars are present outside the central 20'' and are
most likely non-members which move with a large velocity relative to the cluster. The velocity distribution for a central IMBH also clearly deviates from the observed distribution 
at smaller velocities, and a KS test gives a less than $10^{-7}$ chance that both distributions are drawn from the same underlying distribution.
The model with a low black hole retention fraction also does not match the observed distribution of slow moving stars.
We therefore conclude that $\omega$ Cen does not contain an IMBH, or at least, if the cluster contains an IMBH, then its mass must be significantly less than the $47,500$ M$_\odot$
needed to explain the velocity dispersion profile.  Additional simulations with lower IMBH mass fractions and varying stellar-mass black hole fractions would be needed to determine 
the upper mass limit of an IMBH.

\subsubsection{The role of anisotropy and rotation}

\citet{zochietal2017} investigated the influence of orbital anisotropy on the velocity dispersion profile of $\omega$ Cen by fitting {\tt limepy} models \citep{gieleszocchi2015} with
varying degrees of radial anisotropy to the surface and velocity dispersion profile of the cluster. They found that the central velocity dispersion increases 
with increasing orbital radial anisotropy and that {\tt limepy} models could be constructed that reproduced the proper motion dispersion profile of $\omega$ Cen
without the need to invoke an IMBH in the centre of the cluster. While a massive IMBH is already ruled out by the velocity distribution of stars in the centre, 
orbital anisotropy might still have an influence on the parameters of the best-fitting models, especially the amount of stellar mass black holes needed to
reproduce the velocity dispersion profile.

Panel b) of Fig.~\ref{ocen1} compares the orbital anisotropy profile of $\omega$ Cen with the velocity dispersion profile for the best-fitting model with
a 10\% retention fraction of black holes, the best-fitting model with a higher black hole retention fraction and a model with a central IMBH. We define as orbital anisotropy
the ratio of the tangential to the radial velocity dispersion component of the proper motions $\beta=\sigma_t/\sigma_r$. The observed anisotropy $\beta$ is taken
from the measurements of \citet{watkinsetal2015a} and \citet{vanleeuwenetal2000}. In addition, we determine the velocity anisotropy in the outer parts of $\omega$ Cen 
from the {\tt Gaia} DR2 proper motions. The velocity distribution of stars in $\omega$ Cen is
isotropic in the centre out to about 100'', slightly radially anisotropic with $\beta=0.9$ at intermediate radii, before becoming more or less isotropic again 
in the outermost parts. The amount of anisotropy is overall rather small, with the tangential velocity dispersion $\sigma_t$ never differing by more than 10\% 
from the radial velocity dispersion $\sigma_r$. 

In the simulated clusters the velocity profile is also isotropic in the centre before becoming increasingly anisotropic beyond 100'' due to stars being scattered out of the centre onto radial orbits. The simulated clusters
provide an acceptable fit to the anisotropy profile in the inner parts but, except for the IMBH model which is isotropic out to about 1000'', are too anisotropic beyond about 400''. The increasing anisotropy 
is most likely due to the fact that the simulated clusters are isolated, while the Galactic tidal field
deflects stars on their orbits inside $\omega$ Cen and keeps the cluster isotropic. This mismatch could be the reason why the velocity dispersion in the simulated clusters is below the observed velocity
dispersion in the outermost parts of $\omega$ Cen. However, given the good match in the centre, it seems quite unlikely that velocity anisotropy has a significant effect on our results.

In addition to anisotropy, rotation could also influence the results of our fitting since it re-distributes kinetic energy between different spatial directions while the models that we fit to
$\omega$ Cen are non-rotating.
\citet{kamannetal2018} found a rotation amplitude of about 4 km/sec in the central parts of $\omega$ Cen from MUSE spectroscopy. Similarly,
\citet{sollimaetal2019} found a rotational amplitude of $A=4.27 \pm 0.52$ km/sec and a 100\% probability that $\omega$ Cen is rotating by analysing
the {\tt Gaia} DR2 proper motions and stellar radial velocities of \citet{baumgardtetal2019}.
Both values are significantly smaller than the central velocity dispersion, meaning that the influence of rotation is small in the centre
It is therefore also unlikely that rotation has a significant influence on the derived black hole mass fraction and the possible presence of an IMBH.
Finally, stellar binaries could also affect our velocity dispersion estimates. However the binary fraction in $\omega$ Cen is small, only about 13\% \citep{sollimaetal2007} and,
according to the simulations by \citet{ibataetal2011}, the velocity shift of most of these binaries will be close to zero with only few systems producing velocity shifts larger than 20 km/sec.

\subsubsection{Implication for the initial stellar-mass black hole retention fraction}

In this section we compare the 4.6\% mass fraction in stellar-mass black holes that produced the best fit to the observational data of $\omega$ Cen with the estimated BH mass fraction predicted by stellar
evolution theory. To this end, we set up an initial model of $\omega$ Cen assuming the cluster stars follow a Plummer model with an initial total cluster mass of $M=7 \cdot 10^6$ M$_\odot$ and
initial half-mass radius of $r_h=5$ pc, a \citet{kroupa2001} initial mass function between mass limits of 0.1 and 120 M$_\odot$ for the cluster stars and no primordial mass segregation between
high and low-mass stars.
We then apply the effect of stellar evolution to this model by decreasing the masses of the stars and turning the more massive stars into compact remnants and evolve all stars to $T=12$~Gyrs.
We also apply velocity kicks to the stars that turn into black holes.
\begin{figure}
\begin{center}
\includegraphics[width=\columnwidth]{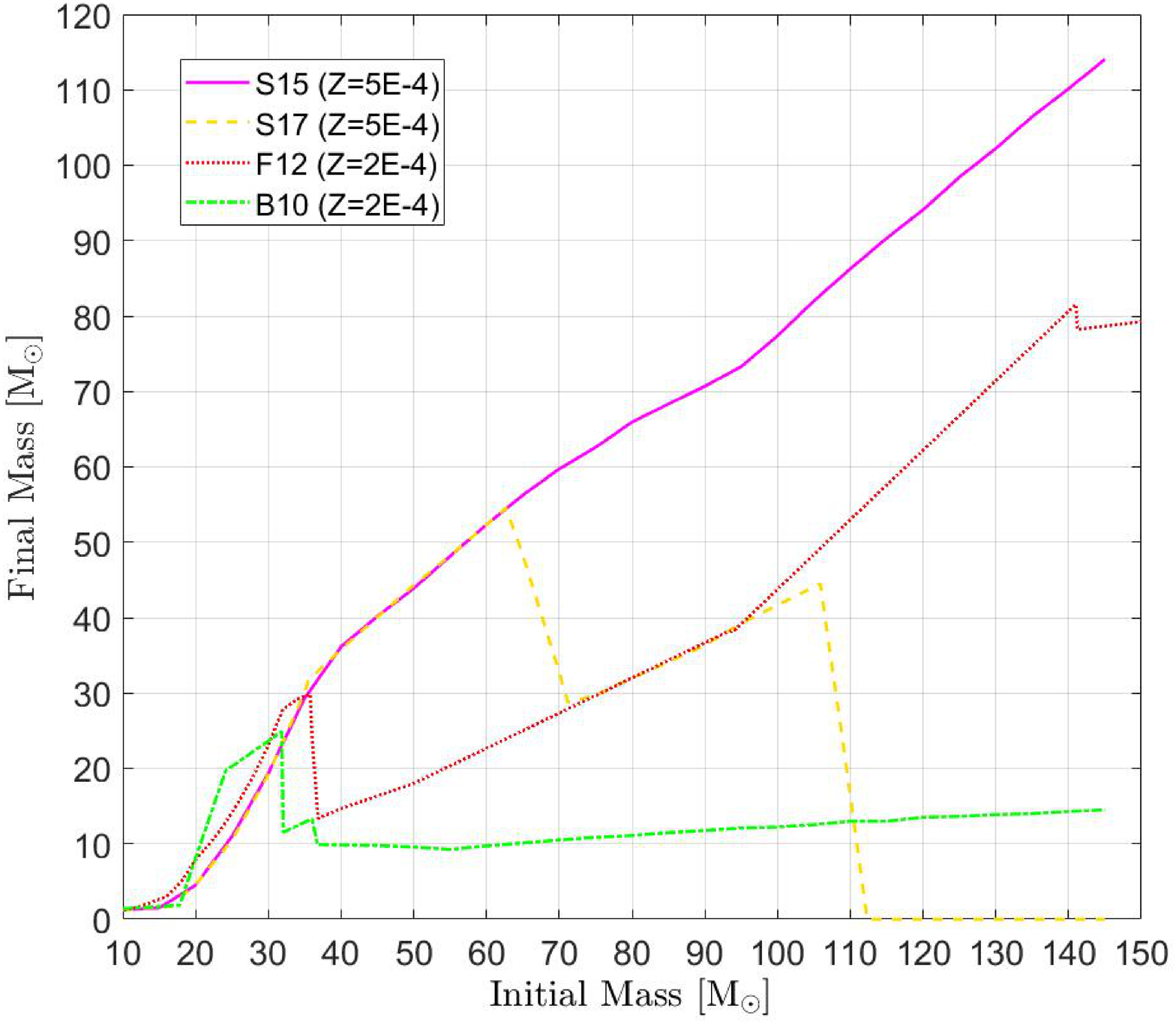}
\end{center}
\caption{Final mass of a black hole as a function of the initial stellar mass for the stellar evolution models of Belczinski et al. (2010) (B10), Fryer et al. (2012) (F12), Spera et al. (2015) (S15) and Spera et al. (2017) (S17).
It can be seen that there is a significant variation between the four models especially for stars with mass $m>30$ M$_\odot$.}
\label{initialfinalmass}
\end{figure}

We assume that stars with masses less than 0.8 M$_\odot$ do not undergo stellar evolution and keep their initial masses.
Stars with initial masses between $0.8 < m < 8$ M$_\odot$ are assumed to be transformed into white dwarfs and we use the initial-final mass function
of \citet{kaliraietal2008} to predict their masses. For the neutron stars we assume that 90\% are removed due to natal kicks, in agreement with the assumption in 
the $N$-body models and that the mass of each neutron star is $m_{NS}=1.3$ M$_\odot$. For stellar-mass black holes, we test four different initial-final mass relations from 
the literature: \citet[][B10]{belczynskietal2010}, \citet[][F12]{fryeretal2012}, \citet[][S15]{speraetal2015} and 
\citet[][S17]{speramapelli2017}. For the S15 and S17 models we assume a cluster metallicity of $Z=0.0005$, close to the average metallicity of $\omega$ Cen according to
\citet{harris1996} and \citet{johnsonpilachowski2010}, while for the B10 and F12 models we assume a metallicity of $Z=0.0002$, which is the metallicity closest to the metallicity of $\omega$ Cen that was
studied in detail in these papers. Fig.~\ref{initialfinalmass} depicts the final mass of a black hole vs. the initial mass of a star for the four stellar evolution models
and the metallicities chosen. We note that the initial-final mass relation for black holes in our $N$-body simulations is given by \citet{belczynskietal2002}, which is similar
to the B10 and (for low stellar masses) F12 models, so these models can most easily be compared with the results of our simulations.

We also apply velocity kicks to the black holes. Following \citet{fryeretal2012}, we assume that the 1D kick velocity $v_{kick}$ is given by the following formula:
\begin{equation}
 v_{kick} = \left(1-f_{fb} \right) \sigma \;\; .
\end{equation}
Here $f_{fb}$ is the mass fraction of the stellar envelope falling back onto the black hole \citep{fryeretal2012}. We assume that $\sigma$, the 1D kick velocity in case of no 
mass fallback, follows a Gaussian distribution and we vary it between $0<\sigma<400$ km/sec to explore the influence of $\sigma$ on our results. We calculate three kick velocities 
for each spatial direction and add them to the velocity of the progenitor star upon formation of a black hole.
Black holes are assumed to escape if their total energy is larger than zero and we sum up the masses of all remaining black holes to obtain the number and mass 
fraction of all black holes after their formation. Since the best-fitting $N$-body model of $\omega$ Cen loses about 1/3 of all black holes between the time
of their formation and T=12 Gyr due to dynamical encounters between single black holes and black hole binaries in the core of the cluster, we finally reduce 
the mass in stellar-mass black holes that we derive from the different stellar-evolution models by 1/3 in order to account for the dynamical mass loss.
\begin{figure}
\begin{center}
\includegraphics[width=\columnwidth]{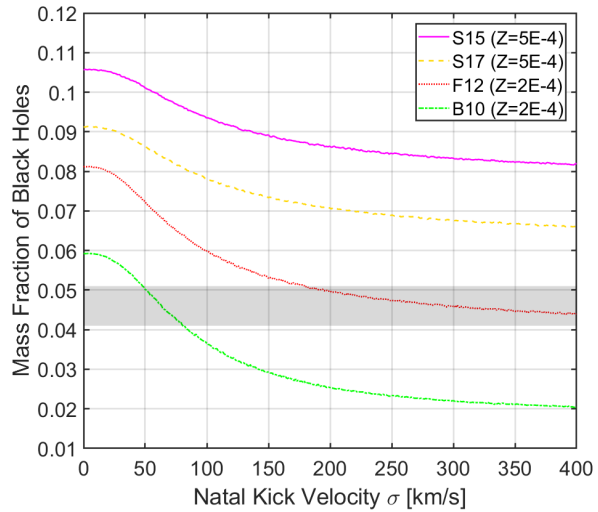}
\end{center}
\caption{Black hole mass fraction after T=12 Gyr as a function of the kick velocities $\sigma$ for the four stellar evolution models depicted in Fig.~\ref{initialfinalmass}.
The grey shaded area shows the predicted black hole mass fraction
in $\omega$ Cen from our $N$-body models. The current BH mass fraction is compatible with the Fryer et al. (2012) (F12) stellar evolution models and a 1D kick velocity of 270 km/sec as found
by \citet{repettoetal2012} or the Belczinski et al. (2010) (B10) models for low kick velocities.}
\label{bhretfrac}
\end{figure}

Fig.~\ref{bhretfrac} depicts the mass fraction in stellar-mass black holes after 12 Gyr for the four different stellar evolution models. The black hole mass fraction 
decreases with increasing kick velocity but becomes roughly constant beyond 300 km/sec due to black holes forming from stars with $f_{fb} \approx 1$ which receive
only small kicks. The grey shaded area shows the observed fraction of $4.6 \pm 0.5 \%$. It can be seen that the observed mass fraction is
compatible with the Belczinski et al. (2010) models for kick velocities up to about 80 km/sec and with the Fryer et al. (2012) models for larger kick velocities.
The Fryer et al. (2012) models produce the right black hole mass fraction for the 1D kick velocity of $\sigma=270$ km/sec found by \citet{repettoetal2012} from observations
of Galactic black holes.
The \citet{speraetal2015} and \citet{speramapelli2017} models overpredict the mass fraction of black holes due to the fact that more massive black holes form in these
models, especially for high mass stars. There are however several ways which could also bring these models into agreement with the observations, for example
a steepening of the high mass end of the initial mass function. In addition, the larger number of massive black holes that are produced in these models could
lead to a more efficient dynamical ejection of stellar-mass black holes, which could help to bring these models into better agreement with the observations.
We therefore conclude that the mass fraction of stellar-mass black holes that
we found from the $N$-body models is, within the model uncertainties, in agreement with the expected fraction based on recent models for the evolution of massive stars. 

\subsection{NGC 6624}

\citet{peutenetal2014} analyzed 16 years of timing data of the low-mass X-ray binary (LMXB) 4U 1820-30 that is located close to the centre of NGC 6624 and found
that this LMXB has a strongly negative period derivative. They suggested that an acceleration of the star along the line-of-sight due to either a dark concentration 
of remnants or an intermediate-mass black hole could be responsible for creating this period change.
Furthermore, \citet{pereraetal2017a} analysed timing observations of the three innermost pulsars in NGC~6624 and concluded
that a 60,000 M$_\odot$ IMBH (later revised down to 20,000 M$_\odot$ by \citet{pereraetal2017b}) is required if their period changes 
are due to a central intermediate mass black hole.
\citet{gielesetal2018} on the other hand  noticed that the strong negative period derivative of 4U 1820-30 is not exceptional when compared to field LMXBs,
making it likely that other explanations like mass-loss from the companion star or spin-orbit coupling can also explain the period change.
They also showed through fitting of the observed surface brightness and velocity dispersion profile of NGC~6624  by multi-mass models that
models without an IMBH are sufficient to explain the observed acceleration of pulsar A in NGC~6624. 
\begin{figure*}
\begin{center}
\includegraphics[width=0.95\textwidth]{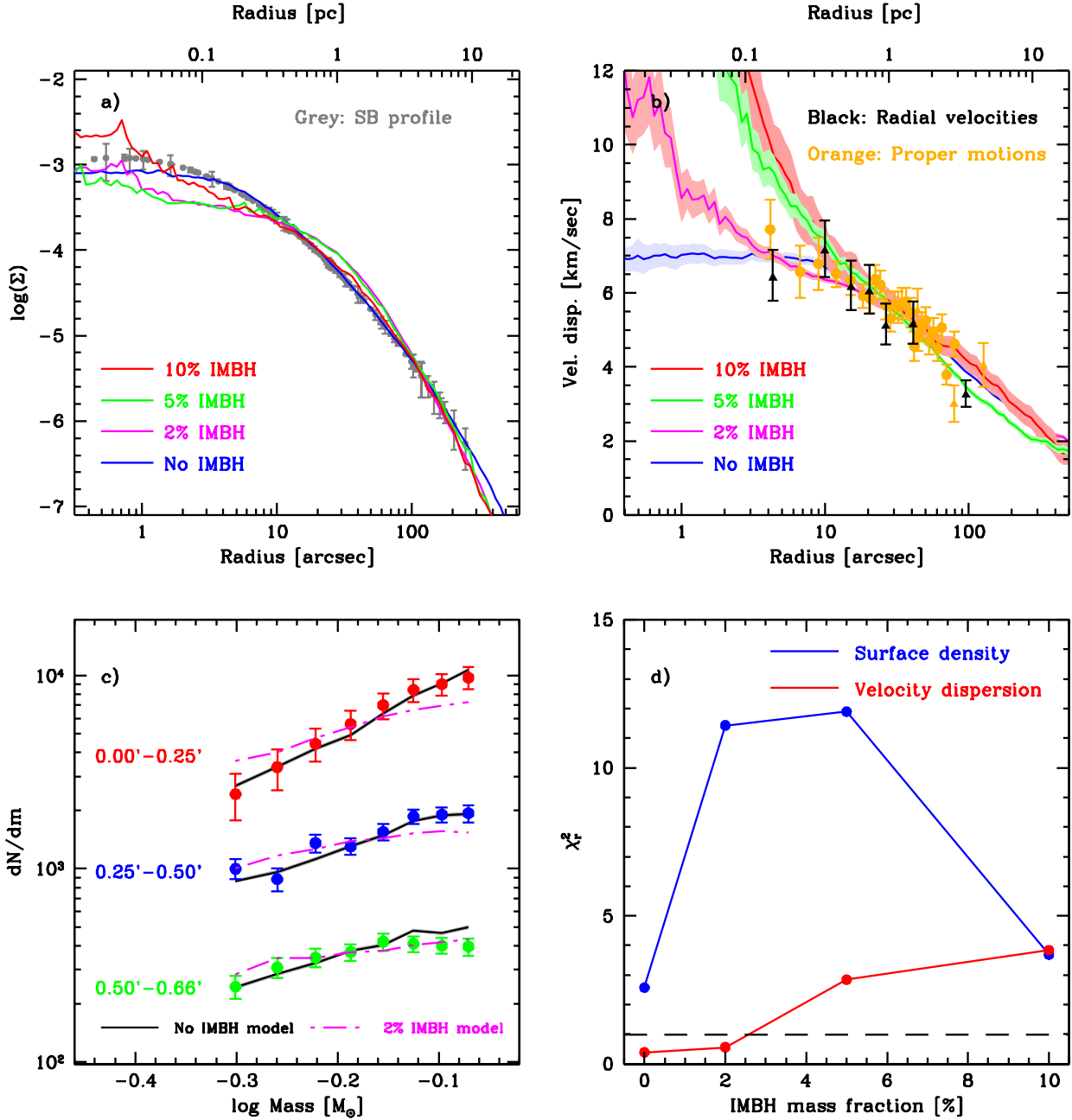}
\end{center}
\caption{Surface density profiles (panel a), velocity dispersion profiles (panel b) and stellar mass functions (panel c) of the best-fitting cluster models with and without an intermediate-mass
black hole (blue lines) and the observed profiles of NGC 6624. Panel c) shows the measured mass functions at three different radii. Only
the comparison with the no-IMBH model is shown here for clarity. Panel d) depicts the reduced $\chi^2_r$ values for the fits of the different models against surface brightness
and velocity dispersion profiles. While the cluster model without
an IMBH is in good agreement with the observed cluster, none of the IMBH models provides a simultaneous fit of the observed surface brightness and velocity dispersion profile.}
\label{ngc6624fits}
\end{figure*}

For a total cluster mass of $55,400 \pm 1,500$ M$_\odot$ \citep{baumgardthilker2018,baumgardtetal2019} even the lower IMBH mass of 20,000 M$_\odot$ 
of \citet{pereraetal2017b} would still imply
that the IMBH would contain more than 1/3 of the total mass of NGC~6624, making NGC~6624 one of the most black hole dominated stellar systems known.
We therefore also fitted our grid of $N$-body simulations to NGC~6624.
Fig.~\ref{ngc6624fits} depicts cluster fits with various IMBH mass fractions to the observed velocity dispersion and surface brightness profile of NGC~6624.
We performed one set of simulations of clusters without a central black hole, and three sets of simulations of star clusters containing central black holes containing
2\%, 5\% and 10\% of the total cluster mass. For each IMBH mass fraction, we searched for the model that produced the best fit to the observed surface brightness
and velocity dispersion profile of NGC~6624. It can be seen that the model without an IMBH is in excellent agreement with the observations since it fits the observed
surface brightness and velocity dispersion profile. There is a slight discrepancy with the observed surface density profile in the innermost few arcsec, however
this discrepancy is nowhere larger than a factor of two and is probably within the uncertainties with which the density profile can be determined in the center.

Our $N$-body simulations therefore confirm earlier results by \citet{gielesetal2018} who also found that no IMBH is required to explain
the observed surface brightness and velocity dispersion profile of NGC~6624.
Models with IMBHs containing less than 5\% of the cluster mass in the form of an IMBH also provide
acceptable fits to the velocity dispersion profile. However, more massive IMBH models start to overpredict the central velocity dispersion. Although we were not able
to run models with IMBHs containing more than 10\% of the cluster mass in the form of an IMBH due to stability issues in the simulation, it is clear from Fig.~\ref{ngc6624fits}
that such models would be excluded even more strongly. In addition all IMBH models produce weak cusps in the cluster centre which are in disagreement with the observed
surface brightness profile. Finally, the IMBH models produce a smaller amount of mass segregation among the cluster stars than the no-IMBH model, which is unable to reproduce the strong change
of the observed mass function with radius (see panel c). This reduced amount of mass segregation is in agreement with theoretical expectations \citep{baumgardtetal2004b,gilletal2008} and
further argues against the presence of an IMBH in NGC~6624.
We derive an upper limit for an IMBH in NGC~6624 of at most a few percent (i.e. about 3,000 M$_\odot$) if only the cluster kinematics 
is taken into account. This value drops to around 1,000 M$_\odot$ if the fit of the surface brightness profile is also taken into account. 

Table~\ref{tab:ngc6624} presents the derived parameters for NGC~6624 from our best-fitting no-IMBH model. The cluster distance was determined by a simultaneous fit of the radial velocity dispersion and proper motion dispersion profiles.
The global mass function slope $\alpha$ (defined as the best-fitting power-law slope $N(m) \sim m^{\alpha}$) was derived for main-sequence stars between 0.2 and 0.8 M$_\odot$.
It's strongly positive value shows that the number of stars is decreasing towards lower masses, meaning that NGC~6624 is highly depleted in low-mass stars. For the best-fitting 
cluster distance, the mass-to-light ratio of NGC~6624 is around 1.4, somewhat below the expected $M/L$ ratio of a cluster with a Kroupa IMF (around 1.7 given the age and metallicity of NGC~6624). This 
can be explained by the highly depleted mass function of NGC~6624. Our values for the total cluster mass, mass-to-light ratio and half-mass radius are in good agreement with those derived
by \citet{gielesetal2018}.
\begin{table}[t]
\caption{Properties of NGC~6624 and observed and predicted accelerations of the milli-second pulsars from our best-fitting no-IMBH model.}
\begin{tabular}{ll}
\hline
Distance & $7425 \pm 273$ pc \\
Mass & $9.40 \pm 0.23 \cdot 10^4$ M$_{\odot}$ \\
M/L ratio & $1.37 \pm 0.17$ M$_\odot$/L$_\odot$ \\
Mass function slope $\alpha$ & +1.5 \\
Relaxation time $T_{RH}$ & $3.0 \cdot 10^8$ yrs \\
Core radius & 0.25 pc \\
Half-mass radius & 2.50 pc \\
Central velocity dispersion & 7.1 km/sec \\
Half-mass density & $7.3 \cdot 10^2$ M$_\odot$/pc$^3$  \\
Central density & $6.1 \cdot 10^5$ M$_\odot$/pc$^3$ \\
\hline
 & \\[-0.2cm]
\multicolumn{2}{l}{Observed period derivatives $\dot{P}/P$}\\[+0.2cm]
PSR B1820-30A$^1$   & $6.22 \cdot 10^{-16}$ s$^{-1}$ \\
PSR B1820-30B$^1$   & $8.32 \cdot 10^{-17}$ s$^{-1}$ \\
PSR J1823-3021C$^1$ & $5.51 \cdot 10^{-16}$ s$^{-1}$ \\
4U 1820-30$^2$ & $-1.7 \cdot 10^{-15}$ s$^{-1}$  \\
 & \\[-0.2cm]
\multicolumn{2}{l}{Maximum accelerations}\\[+0.2cm]
PSR B1820-30A   & $1.03 \cdot 10^{-16}$ s$^{-1}$ \\
PSR B1820-30B   & $1.21 \cdot 10^{-17}$ s$^{-1}$ \\
PSR J1823-3021C & $2.19 \cdot 10^{-17}$ s$^{-1}$ \\
4U 1820-30 & $8.79 \cdot 10^{-17}$ s$^{-1}$ \\
\hline
 & \\[-0.2cm]
\end{tabular}

Notes: 1: from \verb+http://www.naic.edu/~pfreire/GCpsr.html+, 2: from Peuten et al. (2014)
\label{tab:ngc6624}
\end{table}

Table~\ref{tab:ngc6624} also presents the maximum accelerations for the LMXB and the three pulsars with measured period derivatives that are possible in the best-fitting model.
The maximum accelerations were determined for stars at the same projected distance as each of the observed pulsars by varying the distance along the
line of sight until the maximum line-of-sight acceleration was found.
The central density of our best-fitting no-IMBH model is lower than the one found by \citet{gielesetal2018}. As a result, the observed period changes of the pulsars and the LMXB cannot be explained by 
our $N$-body model through an acceleration due to the smooth background cluster potential. Instead, they have to be either due to nearby stars or an internal spin-down of the pulsars. Indeed, the 
period derivative, $\dot{P}$, 
values of pulsars B and C are comparable with observed $\dot{P}$ values of field pulsars of similar period. These pulsars were also not considered by \citet{pereraetal2017b} for the determination of the cluster
potential and IMBH mass. Pulsar A is the most luminous $\gamma$-ray
pulsar known and the observed $\gamma$-ray luminosity requires an intrinsic period derivative that is comparable to the observed value \citep{freireetal2011}.

\begin{figure}
\begin{center}
\includegraphics[width=1.00\columnwidth]{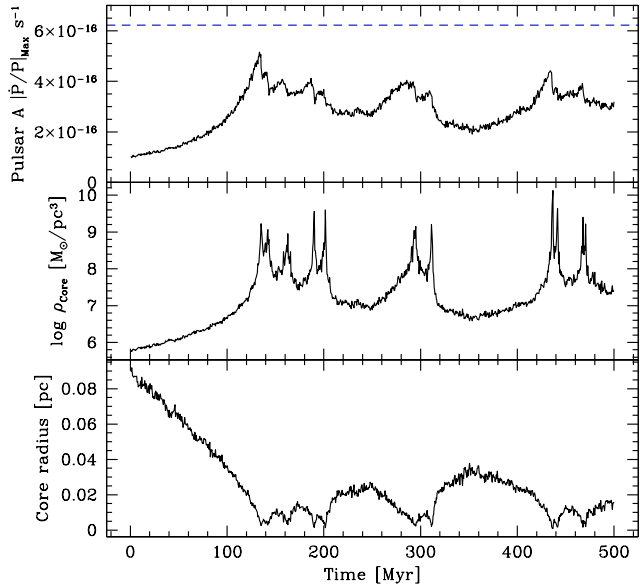}
\end{center}
\caption{Core radius (bottom panel), central density (middle panel) and maximum acceleration of pulsar A (top panel) as a function of time in an $N$-body simulation that uses our best-fitting no-IMBH model
as a starting point. Since NGC~6624 is in core collapse, the core radius and the central density show large fluctuations as the core contracts and re-expands due to the formation of binaries followed by
heating due to encounters between cluster stars and these binaries. However even during the strongest contraction phases the maximum $\dot{P}/P$ value of pulsar A is well below the observed value
(shown by a blue line in the top panel). This implies that most of the observed period change of this pulsar is due to internal processes or a nearby companion, not the background cluster potential.}
\label{ngc6624cc}
\end{figure}

A final caveat to note is that the results of our $N$-body fitting show that NGC~6624 has a half-mass relaxation time of around $10^8$ yrs, much smaller than its age, making it likely that NGC~6624 has 
gone through core-collapse.
Clusters in core-collapse go through core oscillations during which the core continuously collapses and re-expands due to dynamical heating of the core due to binaries formed during
the dense collapse phases \citep{bettwiesersugimoto1984}. This could significantly change the central density without affecting the outer density profile much, i.e. such density fluctuations might not be
visible observationally.
We therefore investigate how the previous results change over time. Fig.~\ref{ngc6624cc} shows the variation of the core radius, central density and maximum acceleration
of pulsar A in an $N$-body simulations that uses our best-fitting no-IMBH model as a starting point. This run was done without stellar evolution, however we do not expect that stellar evolution
will change the results significantly over the 500 Myr timescale depicted in Fig.~\ref{ngc6624cc}. The core radius and central density are calculated according to eq. 2 of \citet{baumgardtetal2002}. 

The model cluster quickly collapses and reaches a central density of around $10^9$ M$_\odot$/pc$^3$, three orders of magnitudes larger than the initial density of the best-fitting $N$-body model, after
about $T=130$ Myr of evolution. We note that the best fit to the surface density profile of NGC~6624 determined by \citet{trageretal1995}, which we use in this paper, is reached at about $T=50$ Myr in this simulation, 
while the surface density profile determined by \citet{gielesetal2018}
corresponds to the density profile reached in the deepest collapse phases. After the initial collapse, core oscillations are clearly visible in the evolution of the core density and the central density 
can fluctuate by about two orders of magnitude within a few 10s of Myr.
However while the core can reach extremely high central densities, it contains only of order 30 stars during the densest collapse phases. As a result the maximum acceleration of a pulsar seen in projection at the same distance
as pulsar~A fluctuates much less and is always a factor two to three below the observed acceleration. We therefore conclude that the
observed period change of pulsar~A must at least in part be due to an internal period change or a nearby companion. It cannot solely be explained by an acceleration due to the general cluster potential, making this pulsar
and also the other two pulsars unsuitable for the determination of the cluster potential and the presence of an IMBH in NGC~6624.

\section{Conclusions}

We have fitted results of dynamical $N$-body simulations to the surface brightness and velocity dispersion profiles of $\omega$ Cen and NGC~6624, two Galactic globular clusters that have been claimed to harbor intermediate-mass black holes (IMBHs).
Our results show that, while IMBH models can be constructed that produce a simultaneous fit to the velocity and surface brightness profile of $\omega$ Cen, these models predict too many fast moving stars within the
central 20'' of the cluster and can therefore be rejected. Instead, we find that a model containing $4.6\%$ of the cluster mass in a centrally concentrated cluster of stellar mass black holes is a viable alternative 
to an IMBH model. This confirms earlier results by \citet{zocchietal2019}. Such a model not only provides a very good fit to the velocity dispersion profile of $\omega$ Cen, but also correctly predicts the velocity distribution 
of stars in the central 20'' of $\omega$ Cen in the HST proper motion survey of \citet{bellinietal2017}. We find that a mass fraction of 4.6\% in stellar mass black hole is compatible with the expected mass fraction due to stellar evolution of massive stars. Our $N$-body simulations show that this centrally concentrated cluster of black holes can have formed due to dynamical mass segregation and energy partition from an initially unsegregated distribution
of stars.

For NGC~6624 we find that a model without an IMBH produces
an excellent fit to the observed surface brightness and velocity dispersion profile as well as the stellar mass function of the cluster, corroborating earlier results by \citet{gielesetal2018}. 
If an IMBH is present at all in this cluster, it must be less
massive than 1,000~M$_\odot$ since more massive IMBHs produce clusters that are in conflict with the observed surface brightness and velocity dispersion profile as well as the amount of mass segregation of NGC 6624. In particular 
IMBHs with more than 5\% of the cluster mass
(corresponding to more than about 3,000 M$_\odot$), produce a strong central rise in velocity dispersion which is neither seen in the HST proper motion dispersion profile of \citet{watkinsetal2015a} nor our radial velocity
dispersion profile. Hence, the observed period derivatives
of the millisecond pulsars in the centre of NGC~6624 cannot be due to an acceleration produced by the smooth background potential of the cluster. Our results therefore show that caution has to be applied when using millisecond pulsars
as probes of globular cluster potentials.

\section*{Acknowledgments}

We thank Andrea Bellini for useful discussions concerning the analysis of the HST proper motions of $\omega$ Cen. We also thank Mark Gieles and an anonymous referee for comments that helped improve the
presentation of the paper. C.U. and S.K. gratefully acknowledge financial support from the European Research Council (ERC-CoG-646928, Multi-Pop).
This paper includes data that has been provided by AAO Data Central (datacentral.org.au). Part of this work is based on data acquired through the Australian Astronomical Observatory, [under program A/2013B/012]. 
Parts of this research were supported by the Australian Research Council Centre of Excellence for All Sky Astrophysics in 3 Dimensions (ASTRO 3D), through project number CE170100013 .
Some of the data presented herein were obtained at the W. M. Keck Observatory, which is operated as a scientific partnership among the California Institute of Technology, the University of California and the National Aeronautics and Space Administration. The Observatory was made possible by the generous financial support of the W. M. Keck Foundation.
The authors wish to recognize and acknowledge the very significant cultural role and reverence that the summit of Maunakea has always had within the indigenous Hawaiian community.  We are most fortunate to have the opportunity to conduct observations from this mountain.


\bibliographystyle{mn2e}
\bibliography{mybib}

\appendix

\section{Individual radial velocities of stars in NGC~6624}

\begin{table*}
\caption{{\tt DEIMOS} stellar radial velocities for stars in the field of NGC 6624. The table gives the 2MASS ID, the right ascension and declination, the average heliocentric radial velocity and its 1$\sigma$ error, the distance from the cluster centre, the 2MASS $J$ and $K_S$ band magnitudes, the membership probability based on the radial velocity and the number of radial velocity measurements. For stars with multiple radial velocity measurements, the probability that the star has a constant radial velocity is given in the final column.  A full version of this Table is available online.}
\begin{tabular}{c@{$\;\;\;$}c@{$\;\;\;$}c@{$\;\;\;$}rrcc@{$\;\;\;$}c@{$\;\;\;$}c@{$\;\;\;$}l}
\hline
 & \\[-0.3cm]
 \multirow{2}{*}{2MASS ID} & RA & DEC & \multicolumn{1}{c}{$R_V$} & \multicolumn{1}{c}{$d$} & $J$ & $K_S$ & Prob. & \multirow{2}{*}{$N_{RV}$} & Prob.\\ 
 & [J2000] & [J2000] & \multicolumn{1}{c}{[km/sec]} & \multicolumn{1}{c}{['']} & [mag] & [mag] & Mem. & & Single\\  
\hline
 & \\[-0.3cm]
 18225355-3023248 &  275.723126 &  -30.390244 &   44.45 $\pm$ 2.95 & 616.74  &  13.95 $\pm$   0.05 &  13.08 $\pm$   0.06 & 0.106 & 1 &  \\
 18225417-3022408 &  275.725745 &  -30.378021 &   -9.71 $\pm$ 3.41 & 602.72  &  11.51 $\pm$   0.02 &  10.39 $\pm$   0.03 & 0.000 & 1 &  \\
 18225559-3020568 &  275.731659 &  -30.349119 &   -4.32 $\pm$ 4.30 & 582.91  &  12.78 $\pm$   0.02 &  11.87 $\pm$   0.02 & 0.000 & 1 &  \\
 18225678-3021341 &  275.736596 &  -30.359499 &  -12.76 $\pm$ 2.46 & 565.99  &  11.94 $\pm$   0.03 &  10.98 $\pm$   0.03 & 0.000 & 1 &  \\
 18234108-3022221 &  275.921187 &  -30.372812 &   50.58 $\pm$ 2.68 &  43.07  &  10.10 $\pm$   0.02 & $\,$  8.90 $\pm$   0.02 & 0.484 & 2 & 0.020\\
 \multicolumn{1}{c}{$\vdots$}  & $\vdots$  & $\vdots$ & \multicolumn{1}{c}{$\vdots$} &\multicolumn{1}{c}{$\vdots$} & \multicolumn{1}{c}{$\vdots$} & $\vdots$ & $\vdots$ & $\vdots$ & \\
\hline
\end{tabular}
\label{indveltab}
\end{table*}

\label{lastpage}
\end{document}